\documentclass{article}
\usepackage{euscript,amsmath,amssymb,amsfonts,graphicx,bm,amsthm,mathtools}
  \usepackage{caption}
\usepackage{subcaption}
\usepackage{pdflscape}

\usepackage{array}

\usepackage{cite} % this makes it so numbered citation list like [1,2,3,4,5,6] would be [1-6].

  \usepackage{paralist}
  \usepackage{graphics} 
\usepackage{soul}

\usepackage{latexsym,amssymb,enumerate,amsmath,amsthm} 
  \usepackage{graphicx} 
  \usepackage{tikz}
     \usepackage[colorlinks=false]{hyperref}
 \hypersetup{urlcolor=blue, citecolor=red}
\usepackage{latexsym,amssymb,enumerate,amsmath} 
\usepackage{pgfplots}
\pgfplotsset{compat=1.18}
\usepackage{soul,xcolor}

\usepackage{lineno}
\newcommand*\patchAmsMathEnvironmentForLineno[1]{%
  \expandafter\let\csname old#1\expandafter\endcsname\csname #1\endcsname
  \expandafter\let\csname oldend#1\expandafter\endcsname\csname end#1\endcsname
  \renewenvironment{#1}%
     {\linenomath\csname old#1\endcsname}%
     {\csname oldend#1\endcsname\endlinenomath}}% 
\newcommand*\patchBothAmsMathEnvironmentsForLineno[1]{%
  \patchAmsMathEnvironmentForLineno{#1}%
  \patchAmsMathEnvironmentForLineno{#1*}}%
\AtBeginDocument{%
\patchBothAmsMathEnvironmentsForLineno{equation}%
\patchBothAmsMathEnvironmentsForLineno{align}%
\patchBothAmsMathEnvironmentsForLineno{flalign}%
\patchBothAmsMathEnvironmentsForLineno{alignat}%
\patchBothAmsMathEnvironmentsForLineno{gather}%
\patchBothAmsMathEnvironmentsForLineno{multline}%
}
\theoremstyle{plain}
\theoremstyle{remark}

\theoremstyle{definition}

\usepackage{graphicx, mathtools,amsmath,float,chemfig,tikz,siunitx}
\def\[#1\]{\begin{align*}#1\end{align*}}
\newcommand{\dd}{\text{d}}
\newcommand{\eps}{\varepsilon}
\renewcommand{\phi}{\varphi}
\newcommand{\ka}{k_{\text{a}}}
\newcommand{\kc}{k_{\text{c}}}
\newcommand{\ki}{k_{\text{i}}}
\newcommand{\kd}{k_{\text{d}}}

\newcommand{\kaa}{k_{\text{a,1}}}
\newcommand{\kca}{k_{\text{c,1}}}
\newcommand{\kab}{k_{\text{a,2}}}
\newcommand{\kcb}{k_{\text{c,2}}}
\newcommand{\mfar}{I}
\newcommand{\cfar}{S}
\newcommand{\cfara}{S_1}
\newcommand{\cfarb}{S_2}
\newcommand{\cinf}{c_{\infty}}
\newcommand{\minf}{m_{\infty}}
\newcommand{\uinf}{u_{\infty}}
\newcommand{\cinfa}{c_{\infty,1}}
\newcommand{\cinfb}{c_{\infty,2}}
\newcommand{\binf}{b_{\infty}}
\newcommand{\binfa}{b_{\infty,1}}
\newcommand{\binfb}{b_{\infty,2}}
\newcommand{\binfc}{b_{\infty,\text{c}}}
\newcommand{\binfm}{b_{\infty,\text{m}}}

\begin{document}

%\preprint{}

\title{Reaction kinetics of membrane receptors: a spatial modeling approach}


\author{An\i l Cengiz\thanks{Department of Mathematics, University of Utah, Salt Lake City, UT 84112 USA.}\and Sean D. Lawley\thanks{Department of Mathematics, University of Utah, Salt Lake City, UT 84112 USA (\texttt{lawley@math.utah.edu}).}
}
\date{\today}
\maketitle

%\newpage
\begin{abstract}
The interactions between diffusing molecules and membrane-bound receptors drive numerous cellular processes. In this work, we develop a spatial model of molecular interactions with membrane receptors by homogenizing the cell membrane and describing the evolution of both molecular diffusion and molecule-receptor interactions. By analyzing a resulting partial differential equation coupled to ordinary differential equations, we derive analytical expressions for the steady-state  molecular influx rate in four prototypical interaction scenarios: Michaelis-Menten kinetics, Substrate Competition, Competitive Inhibition, and Uncompetitive Inhibition. For each scenario, we show how to modify the classical well-mixed reaction rate theory to resolve spatial features inherent to receptors bound to cell membranes. We find that naive well-mixed calculations significantly overestimate reaction rates in certain biophysical parameter regimes.
\end{abstract}

\newpage
\tableofcontents

%%%%%%%%%%%%%%%%%%%%%%%%%%%%%%%%%%%%%%%%%%%%%%%%%%%%%%%%%%%%%%%%%%%%%%%%%%%%%%%%%%%%%%%%%%%%%%%%%%%%%%%%%%%%%%%%%%%%%%%%%%%%%%%%%%%%%%%%%%%%

\newpage
\section{Introduction}
Cellular processes are often initiated and driven by membrane receptors interacting with diffusing molecules. Indeed, a molecule binding to a membrane receptor and/or being transported into the cell is often the first step of a signaling cascade leading to various cellular actions \cite{lauffenburger_1996}. For example, neurotransmission of information is carried out by neural receptors taking in diffusing neurotransmitters released by other neurons \cite{hammond_2015}. Quorum sensing involves membrane receptors binding to diffusing auto-inducers produced by cells to detect cell population density \cite{tommonaro_2019}. Bacterial bio-degradation takes place when bacterial surface receptors bind and transport bio-pollutants into the bacterial cytosol \cite{mohanan_2020}. The rate of such intra-cellular processes depends on the binding kinetics of membrane receptors. In this paper, we study how such kinetics depend on spatial features.

To introduce the problems of interest, we briefly review the century-old work by Michaelis and Menten \cite{michaelis_2013,johnson_2011} and Briggs and Haldane \cite{briggs_1925}. Consider the following substrate-enzyme reaction scheme,
\begin{align}\label{eq:mm}
        \schemestart
S + E \arrow {->[$\ka$]} ES \arrow {->[$\kc$]} E + P.
\schemestop
\end{align}
In words, a substrate S binds at the association rate $\ka$ to an enzyme E to form a complex ES which then produces a product P at catalysis rate $\kc$ (for simplicity, we assume that the first step in \eqref{eq:mm} is irreversible). Assuming that the substrates and enzymes are well-mixed in a three-dimensional domain, the reaction rate (i.e.\ the product formation rate $\dd P/\dd t$) is
\begin{equation}
    {\overline{V}}
    =\frac{{V}_{\max}S}{{K} + S},\label{eq:mm_rex_rate0}\end{equation}
where $S$ is the substrate concentration, and the maximum reaction rate ${V}_{\max}$ and half-saturation constant ${K}$ are given by
\begin{align}\label{eq:mmp}
\begin{split}
    {V}_{\max}
    &=\kc E_{\text{tot}},\\
    {K}
    &=\kc/\ka,
\end{split}
\end{align}
where $E_{\text{tot}}$ is the total enzyme concentration \cite{briggs_1925}. 

Suppose the ``enzymes'' E in \eqref{eq:mm} are receptors embedded on the membranes of a well-mixed concentration $C$ of cells in the domain. Naively applying Michaelis-Menten theory implies that the rate of product formation is \eqref{eq:mm_rex_rate0} with constants ${V}_{\max}$ and ${K}$ in \eqref{eq:mmp} with 
\begin{align*}
    E_{\text{tot}}=NC,
\end{align*}
where $N$ is the number of receptors per cell. Further, if association happens immediately upon contact between substrates and receptors, then diffusion-controlled reaction rate theory (detailed below) suggests setting the association rate to be
\begin{align*}
    \ka
    =4\eps RD,
\end{align*}
assuming $R$ is the radius of each cell, each receptor is a disk of radius $\eps R$ with $0<\eps<1$, and $D$ is the sum of the diffusivities of substrates and cells.

\begin{figure}[t]
\centering
\includegraphics[width=1\textwidth]
{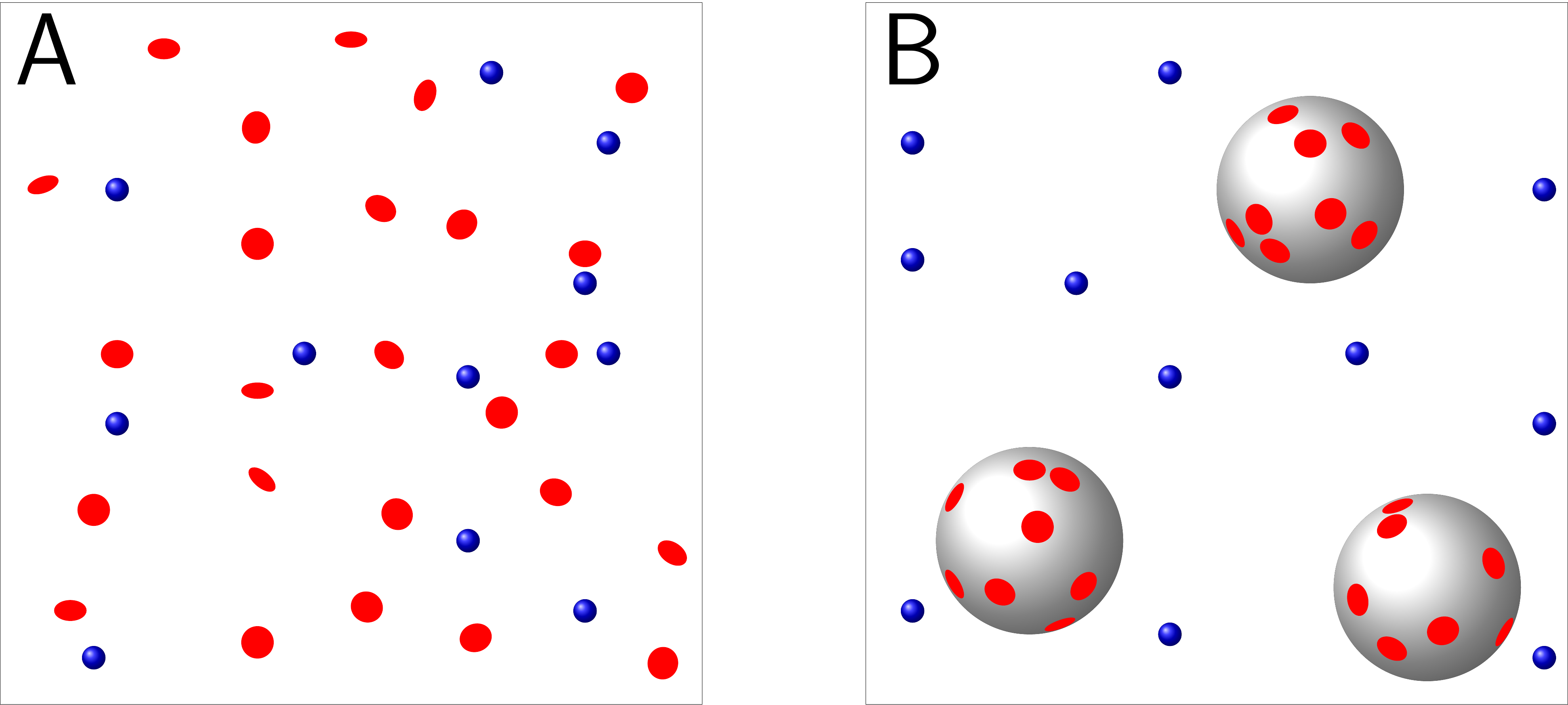}
\caption{\label{fig:schem} Panel A: Substrate molecules (blue spheres) and receptors (red disks) are well-mixed and each receptor freely diffuses. Panel B: Receptors are bound to cell membranes (gray spheres).} 
\end{figure}

How accurate is this naive application of classical Michaelis-Menten theory? The classical theory assumes that the receptors are well-mixed in the domain, i.e.\ receptors are disks diffusing independently in the three-dimensional volume (see Figure~\ref{fig:schem}A). However, in the actual scenario of interest, (i) receptors are embedded on two-dimensional cell membranes, and (ii) receptors on the same cell may be sufficiently close to each other to effectively compete for substrates (see Figure~\ref{fig:schem}B). How can the classical, well-mixed theory be modified to incorporate these spatial features and spatiotemporal correlations? Furthermore, if there are multiple substrates, how accurate is an analogous application of classical Substrate Competition \cite{pocklington_1969} theory? If there are inhibitory agents, how accurate is an application of classical Competitive Inhibition or Uncompetitive Inhibition \cite{dixon_1953} theory? The purpose of this paper is to investigate these questions. 

\begin{figure}
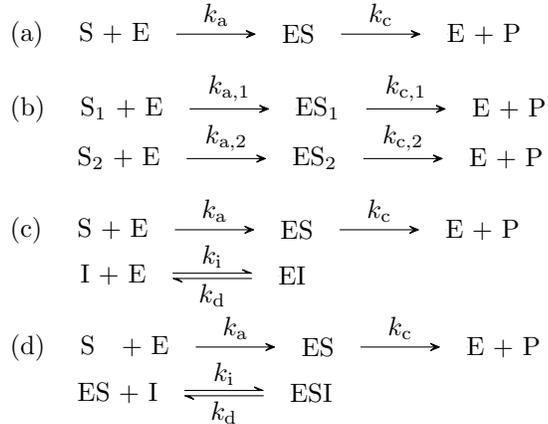

\schemestart
(a) \quad S + E \arrow {->[$\ka$]} ES \arrow {->[$\kc$]} E + P 
\schemestop
\\
\\
\schemestart
(b) \quad \chemfig{S_1} + E \arrow{->[$k_{\textup{a},1}$]} \chemfig{ES_1} \arrow{->[$k_{\textup{c},1}$]} E + \chemfig{P}
\schemestop\\
\schemestart
\phantom{(c)} \quad \chemfig{{S_2}} + E \arrow{->[$k_{\textup{a},2}$]} \chemfig{E{S_2}} \arrow{->[$k_{\textup{c},2}$]} E + \chemfig{P}
\schemestop 
\\
\\
\schemestart 
(c)  \quad S + E \arrow{->[$\ka$]} ES \arrow{->[$\kc$]} E + P 
\schemestop \\
\schemestart
\phantom{(d)} \quad 
I + E \arrow {<=>[$\ki$][$\kd$]} EI 
\schemestop
\\
\\
\schemestart
(d) \quad S\phantom{E} + E \arrow {->[$\ka$]} ES \arrow {->[$\kc$]} E + P 
\schemestop \\
\phantom{(e)} \quad \schemestart 
ES + I \arrow {<=>[$\ki$][$\kd$]} ESI
\schemestop
\caption{\label{fig:schemes} Kinetic schemes for (a) Michaelis-Menten Kinetics, (b) Substrate Competition, (c) Competitive Inhibition, and (d) Uncompetitive Inhibition. S, $\text{S}_1$, and $\text{S}_2$ represent different substrates, E represents membrane receptors, ES, $\text{ES}_1$, and $\text{ES}_2$ represent substrate-receptor complexes, I represents an inhibitor, EI represents a receptor-inhibitor complex, ESI represents a receptor-substrate-inhibitor complex, and P represents products.} 
\end{figure}

We model the reaction kinetics of membrane receptors by incorporating the spatial dynamics of these systems and mathematically describing the molecule-receptor interactions on the membrane. We characterize the rate of molecular influx through receptor proteins in four common molecule-receptor interaction scenarios: Michaelis-Menten Kinetics \cite{michaelis_2013,johnson_2011,briggs_1925}, Substrate Competition \cite{pocklington_1969},  Competitive Inhibition \cite{dixon_1953}, and Uncompetitive Inhibition \cite{dixon_1953}. The kinetic schemes for these interaction scenarios are shown in Figure~\ref{fig:schemes}. 

The Michaelis-Menten kinetics discussed in this work describe fundamental substrate-enzyme interactions that occur on the cell surface. Classic examples include G-protein activation, where diffusing ligands bind to G protein-coupled receptors (GPCRs) \cite{weis_2018, heydenreich_2022}, and the uptake of glucose molecules by permease proteins in \emph{E.~coli} \cite{natarajan_1999, carreon_2023}. In some scenarios, membrane receptors may bind with multiple diffusing species, leading to Substrate Competition for the active site of the membrane-bound enzyme \cite{pocklington_1969}. For instance, in the Mitogen-activated protein kinase (MAPK) pathway, three substrates—Capicua (Cic), Bicoid (Bcd), and Hunchback (Hb)—compete for the same active site on the MAPK enzymes \cite{kim_2011}. Competition at the active site can also involve Competitive Inhibition \cite{dixon_1953}, in which inhibitors block the site and make it unavailable to substrates. This mechanism of action is employed by drugs like statins, which competitively inhibit HMG-CoA reductase to lower cholesterol production \cite{satish_2016} and antiretroviral drugs such as HIV protease inhibitors, which prevent viral replication by blocking the protease active site \cite{wang_2015}. Another important scenario is Uncompetitive Inhibition \cite{dixon_1953}, where an inhibitor binds to the membrane-bound substrate-enzyme complex, prolonging the inactive state and slowing catalysis. In the case of the Alzheimer's drug memantine, this mechanism involves blocking the N-methyl-D-aspartate (NMDA) receptor when it is in its open state, which represents the substrate-enzyme complex formed by the binding of two glutamate and two glycine molecules to the receptor. By binding to this complex, memantine prevents current flow through the channel, thereby reducing neurotoxicity \cite{johnson_2006}.

For each of these four interaction scenarios (see Figure~\ref{fig:schemes}), we derive the molecular influx rate (i.e.\ the reaction rate or product formation rate $V=\dd P/\dd t$) assuming that the receptor protein size is small relative to the cell radius ($\eps\ll1$) and that cells are well-separated in the domain. We show that this spatial molecular influx rate has a similar form to the reaction rate formula of the corresponding non-spatial substrate-enzyme system. The key difference is that the half-saturation parameter is a function of the substrate concentration $S$ for the spatial system. To illustrate for Michaelis-Menten kinetics, the half-saturation constant ${K}$ in \eqref{eq:mm_rex_rate0}-\eqref{eq:mmp} is multiplied by a function $\varphi=\varphi(S/K)$ of the ratio $S/K$ for the spatial system. Furthermore, we show that this function  (i) limits to unity in the high substrate concentration regime,
\begin{align*}
    \lim_{S/K\to\infty}\varphi(S/K)
    =1,
\end{align*}
and (ii) has the following limit in the low substrate concentration regime,
\begin{align}\label{eq:lim2}
    \lim_{S/K\to0}\varphi(S/K)
    =1+\kappa>1,
\end{align}
where $\kappa=\eps N/\pi>0$ describes the trapping rate of the cell surface \cite{shoup_role_1982}. 
We show that \eqref{eq:lim2} implies that the reaction rate $\overline{V}$ for the well-mixed system (i.e.\ for Figure~\ref{fig:schem}A) is a drastic overestimate of the reaction rate $V$ for the spatial system (i.e.\ for Figure~\ref{fig:schem}B) in some parameter regimes of biophysical interest. In particular, we show that
\begin{align*}
    V\ll\overline{V}\quad\text{if }S/K\not\gg \kappa\gg1.
\end{align*}
Hence, we (a) delineate when a naive application of the well-mixed Michaelis-Menten theory does or does not accurately describe the spatial system, and (b) show how to correct the well-mixed theory to account for spatial effects. We obtain similar results for the other three reaction schemes in Figure~\ref{fig:schemes} (Substrate Competition, Competitive Inhibition, and Uncompetitive Inhibition).

The remainder of the paper is organized as follows. In section~\ref{sec:analysis}, we formulate the reaction kinetics problem for Michaelis-Menten kinetics. Our model consists of a partial differential equation (PDE) coupled to an ordinary differential equation (ODE), and we use boundary homogenization theory to derive the reaction rate. Section~\ref{sec:results} presents our findings on the four types of interactions we examined between diffusing substrate molecules and membrane receptors. We provide mathematical expressions for the spatially-resolved reaction rates and compare them to the corresponding well-mixed reaction rates. We conclude by discussing related work and potential extensions. The reaction rate derivations for Substrate Competition, Competitive Inhibition, and Uncompetitive Inhibition are presented in the appendix in section~\ref{sec:appendix}.

%%%%%%%%%%%%%%%%%%%%%%%%%%%%%%%%%%%%%%%%%%%%%%%%%%%%%%%%%%%%%%%%%%%%%%%%%%%%%%%%%%%%%%%%%%%%%%%%%%%%%%%%%%%%%%%%%%%%%%%%%%%%%%%%%%%%%%%%%%%%%%%%%%%%%%%%%%%%%%%%%%%%%%%%%%%%%%%%%%%%%%%%%%%%%%%%%%%%%%
\section{Mathematical model and analysis\label{sec:analysis}}
Interactions between diffusing molecules and receptors can be described by ODEs if molecules and receptors are ``well-mixed'' in the volume of the spatial domain (as in Figure~\ref{fig:schem}A). However, receptors are bound to the two-dimensional membrane, and thus are not homogeneously distributed throughout the three-dimensional domain \cite{hammond_2015} (as in Figure~\ref{fig:schem}B).

To account for these spatial features, we start with a PDE which tracks the diffusion of molecules and their interactions with receptors via mixed boundary conditions at the cell surface. We then employ the theory of boundary homogenization \cite{berg_1977, handy_2021} to homogenize the cell membrane and introduce a coupled PDE-ODE system that describes the molecular diffusive process(es) with PDE(s) in the three-dimensional volume with boundary conditions involving ODEs which describe the interactions between the molecules and membrane receptors \cite{gomez2019linear, gomez2021pattern, bressloff2017dynamically}.

The coupled PDE-ODE system changes based on the interaction scenarios between diffusing molecules and membrane receptors. However, the initial binding rate of the diffusing molecules is independent of the interaction scenario. Therefore, in the next subsection, we first use boundary homogenization to derive the binding rate of diffusing molecules to a heterogeneous boundary without enzyme kinetics in a framework akin to Berg and Purcell's classical model \cite{berg_1977}. We then go on in section~\ref{subsec:analysis} to model the full spatial problem for a specific interaction type, Michaelis-Menten kinetics, between diffusing molecules and membrane receptors. 

%%%%%%%%%%%%%%%%%%%%%%%%%%%%%%%%%%%%%%%%%%%%%%%%%%%%%%%%%%%%%%%%%%%%%%%%%%%%%%%%%%%%
\subsection{Derivation of association rate $\ka$\label{subsec:bound_homog}}

Consider a spherical cell $\Omega$ of radius $R$ centered at the origin with diffusing substrate molecules surrounding it. Denote the substrate concentration at time $t$ by $c(r,\theta,\phi,t)$ with $(r,\theta,\phi)$ as the spherical coordinates. Assume that the cell has $N\gg 1$ absorbing receptor proteins $\{\Omega_1, \dots, \Omega_{N}\}$ that are locally circular regions of radius $\eps R$ with $\eps\ll 1$ and are roughly evenly distributed on an otherwise reflecting cell surface. The molecular concentration $c$ then satisfies the following PDE with mixed Dirichlet-Neumann boundary conditions,
\begin{equation*}
\label{eq:het_pde}
\begin{alignedat}{2}
\partial_t c &=D \Delta c, \quad &&r>R, \\
c&=0, \quad &&(r,\theta,\phi) \in \{\Omega_1,\dots,\Omega_N\}, \\
D\partial_r c&=0, \quad &&(r,\theta,\phi) \in \Omega\setminus\{\Omega_1,\dots,\Omega_N\},
\end{alignedat}
\end{equation*}
where $D$ is the diffusivity of the molecular compound. Assuming that the receptors occupy a small fraction of the cell surface,
\begin{align*}
    \frac{N\eps^2}{4}\ll1,
\end{align*}
the method of boundary homogenization \cite{berg_1977} approximates the mixed boundary condition PDE above by the following PDE with a homogeneous boundary condition at the cell surface,
\begin{equation}
\label{eq:bound_homog_pde}
\begin{alignedat}{2}
\partial_t c &= D \Delta c, \quad &&r>R, \\
D\partial_r c&=\kappa\frac{D}{R}c, \quad &&r=R,
\end{alignedat}
\end{equation}
where 
\begin{align*}
    \kappa
    =\frac{\eps N}{\pi}
\end{align*}
is the dimensionless trapping rate \cite{shoup_role_1982}.

Going forward, we work with the homogeneous boundary in \eqref{eq:bound_homog_pde}. Adopting a similar approach to that in \cite{handy_2021}, we consider the following kinetic scheme,
\begin{align}\label{eq:association}
    \schemestart
S + E \arrow{->[$\ka$]} E + P,
\schemestop
\end{align}
and derive an expression for the association rate $\ka$. The scheme \eqref{eq:association} is equivalent to the Michaelis-Menten scheme in \eqref{eq:mm} except the catalysis rate is taken to be infinite (i.e.\ $\kc=\infty$) so that receptors instantly absorb bound molecules and are immediately available for binding additional molecules. Since the number of available receptors on the two-dimensional cell membrane does not change in this scheme, the receptor concentration $u(t)$ is constant in time and given by the number of receptors per cell surface area,
\begin{align*}
    u(t)
    =u_0
    :=\frac{N}{4\pi R^2}.
\end{align*}
We note that $c$ is the concentration of molecules in the three-dimensional volume and thus has units of number per unit volume, whereas $u$ is the concentration of available receptors on the two-dimensional cell surface and thus has units of number per unit area.

By the law of mass action, the change in the concentration of diffusing molecules at the cell boundary satisfies
\begin{align}
\label{eq:bound_homog_kinetic_bc}
D\partial_r c = \ka c u = \ka c u_0, \quad r=R.
\end{align}
In order for the homogenized boundary condition in \eqref{eq:bound_homog_pde} to be equivalent to the boundary condition from the kinetic scheme in \eqref{eq:bound_homog_kinetic_bc}, the association rate $\ka$ must be 
\begin{align}\label{eq:ka}
    \ka= 4\eps RD.
\end{align}

%%%%%%%%%%%%%%%%%%%%%%%%%%%%%%%%%%%%%%%%%%%%%%%%%%%%%%%%%%%%%%%%%%%%%%%%%%%%%%%%%%%%
\subsection{Uptake derivation for Michaelis-Menten kinetics \label{subsec:analysis}}

We now analyze the kinetics of membrane receptors with Michaelis-Menten kinetics shown in Figure~\ref{fig:schemes}a and derive the reaction rate for such systems. The derivation for other kinds of interaction scenarios is similar and is presented in the appendix in section~\ref{sec:appendix}.  

As above, consider a spherical cell $\Omega$ of radius $R$ centered at the origin with diffusing molecules surrounding it and denote their concentration by $c=c(r,\theta,\phi,t)$. Assume that the concentration of diffusing molecules away from the cell is fixed at $\cfar>0$. The concentration $c$ then satisfies the following diffusion PDE and the far-field condition,
\begin{align}
\begin{split}\partial_t c &= D \Delta c, \quad r>R, \\
\lim_{r\to\infty} c&=\cfar>0,\end{split} \label{eq:mm_bind_pde}\end{align}
where $D$ is the diffusivity of the molecular compound. The far-field condition in \eqref{eq:mm_bind_pde} reflects our assumption that different cells are sufficiently well-separated that they can be treated as non-interacting. The boundary condition at the cell surface is
\begin{align}D\partial_r c =\ka uc, \quad r=R,\label{eq:mm_bc}\end{align}
where the concentration of available receptors and receptor-substrate complexes at time $t$ are respectively denoted by $u(t)$ and  $b(t)$ and governed by the following ODEs,
\[\frac{\dd}{\dd t}u&=-\ka c(R)u + \kc b, \\
\frac{\dd}{\dd t}b&=\ka c(R)u-\kc b, \]
with $c(R):=c(R,\theta,\phi,t)$ for notational ease. In this system, receptors can be in one of two states: available or part of a receptor-substrate complex. Therefore, the total receptor concentration in this system is conserved. Since we have $N$ receptors, the total receptor concentration is
\begin{align*}
    u_0:=u(t)+b(t)=\frac{N}{4\pi R^2}.
\end{align*}
This conservation law can then be used to reduce the system of two ODEs to a single ODE and create the following linear algebraic equation that describes the steady-state behavior of this system, 
\[0&=-\ka \cinf(R)(u_0-\binf)+\kc \binf, \]
where $\cinf,\binf$ denote the steady-state concentration of the diffusing molecules and the receptor-substrate complexes, respectively.  The solution to this linear system is
\[\binf=\frac{\ka \cinf(R) u_0}{\ka \cinf(R) + \kc }.\]

In steady-state, the solution to the PDE in \eqref{eq:mm_bind_pde} evaluated at $r=R$ is given by
\begin{align}\cinf(R)=\cfar\left(1-a \right) \label{eq:cinf_mm},\end{align}
with $a$ as an undetermined integration constant. Substituting the expression for $\cinf(R)$ into the boundary condition in \eqref{eq:mm_bc} and recalling from \eqref{eq:ka} that $\ka=4\eps DR$, the constant $a$ can be shown to be a root of the following quadratic polynomial in $a$,
\begin{align}
A_2 a^2 + A_1 a + A_0,\label{eq:mm_poly}
\end{align}
if
\begin{align}\label{eq:As}
\begin{split}
    A_2
    &=\cfar/{K},\\
    %=4\eps DR\cfar, \\
    A_1
    &=
    -\big(\kappa + \cfar/{K} +1 \big)
    ,\\
    A_0
    &=\kappa,
\end{split}
\end{align}
and ${K}=\kc/\ka$. 
Using the quadratic formula and the condition that $0< a <1$ due to the steady-state concentration in \eqref{eq:cinf_mm} being positive, the root of \eqref{eq:mm_poly} is 
\begin{align}
a
=\frac{-A_1 - \sqrt{A_1^2 - 4 A_2A_0}}{2A_2}=\frac{-2A_0}{A_1-\sqrt{A_1^2-4A_2A_0}}. \label{eq:quad_form}
\end{align}
We compute the reaction rate $V$ by multiplying the cell concentration $C$ by the molecular flux into a single cell,
\begin{align}\label{eq:Va}
    V&:=C\int_{\Omega} D \nabla c \cdot \, d\sigma%,\\
    %&=4\pi R D \cfar a, \\
    =C4\pi R D \cfar a.
\end{align}
Using \eqref{eq:As}-\eqref{eq:quad_form}, the reaction rate $V$ in \eqref{eq:Va} can be algebraically manipulated to have the form in \eqref{eq:mmspatial}.

%%%%%%%%%%%%%%%%%%%%%%%%%%%%%%%%%%%%%%%%%%%%%%%%%%%%%%%%%%%%%%%%%%%%%%%%%%%%%%%%%%%%%%%%%%%%%%%%%%%%%%%%%%%%%%%%%%%%%%%%%%%%%%%%%%%%%%%%%%%%%%%%%%%%%%%%%%%%%%%%%%%%%%%%%%%%%%%%%%%%%%%%%%%%%%%%%%%%%%%%%%%%%%%%%%%%%%%%%%%%%%%%%%%%%%%%%%%%%%%%%%%%%%%%%%%%%%%%%%%%%%%%%%%%%%%%%%%%%%%%%%%%%%%%
\section{Results\label{sec:results}}

The analysis in subsection~\ref{subsec:analysis} for Michaelis-Menten binding is carried out, as detailed in the appendix in  section~\ref{sec:appendix}, to derive the reaction rate $V$ in the remaining three interaction scenarios: Substrate Competition, Competitive Inhibition, and Uncompetitive Inhibition. For each interaction scenario, we derive an analytical expression for $V$ as a function of the various biophysical parameters. We now present these reaction rates and compare them to the reaction rate of the corresponding well-mixed substrate-enzyme system.

%%%%%%%%%%%%%%%%%%%%%%%%%%%%%%%%%%%%%%%%%%%%%%%%%%%%%%%%%%%%%%%%%%%%%%%%%%%%%%%%%%%%
\subsection{Michaelis-Menten kinetics for membrane-bound receptors\label{sec:mm}}

The well-mixed reaction rate $\overline{V}$ for the following Michaelis-Menten substrate-enzyme kinetic scheme,
\begin{align}\label{eq:mm000}
            \schemestart
S + E \arrow {->[$\ka$]} ES \arrow {->[$\kc$]} E + P,
\schemestop
\end{align}
is \cite{briggs_1925}
\begin{equation}\label{eq:mmwm}
    {\overline{V}}
    =\frac{{V}_{\max}S}{{K} + S},
    \end{equation}
where $S$ is the substrate concentration and
\begin{align}
\begin{split}\label{eq:mmend}
    {V}_{\max}
    &=\kc E_{\text{tot}}
    =\kc NC,\\
    {K}
    &=\kc/\ka,
\end{split}
\end{align}
where $E_{\text{tot}}=NC$ is the total enzyme/receptor concentration.

\begin{figure}[t]
\centering
\includegraphics[width=1\textwidth]
{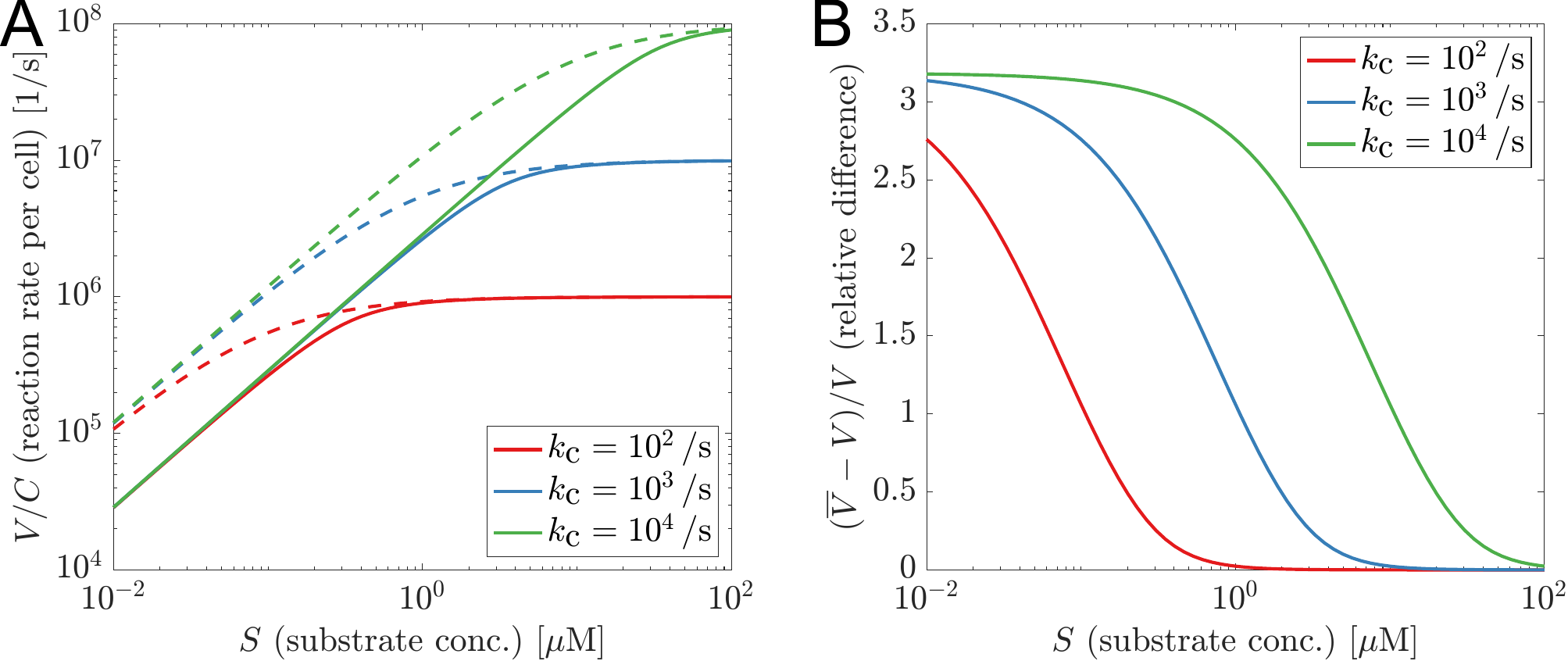}
\caption{\label{fig:mm} Michaelis-Menten kinetics. Panel A: The solid curves show the spatial reaction rate ($V/C$ where $V$ is in \eqref{eq:mmspatial}) as a function of the substrate concentration $S$ for different values of the catalysis rate $\kc$. The dashed curves show the well-mixed reaction rate ($\overline{V}/C$ where $\overline{V}$ is in \eqref{eq:mmwm}). Panel B: The relative difference between the spatial reaction rate $V$ and the well-mixed reaction rate $\overline{V}$ increases at low substrate concentrations $S$ and fast catalysis rates $\kc$. Parameters are given in section~\ref{sec:mm} for both panels.} 
\end{figure}

In section~\ref{sec:analysis}, we derived the following reaction rate for the corresponding spatial system,
\begin{align}\label{eq:mmspatial}
    V
    =\frac{V_{\max}S}{\varphi(S/{K}){K}+S},
\end{align}
where $\ka$ is in \eqref{eq:ka} and $\varphi(x)$ is the following dimensionless function,
\begin{align}\label{eq:varphi}
    \varphi(x)
    &=\frac{1}{2}\Big(1+\kappa-x+\sqrt{\big(1+\kappa+x\big)^2-4\kappa x}\Big),
\end{align}
where $\kappa=\eps N/\pi>0$ describes the effective trapping rate of the cell surface. We note that $\varphi(x)\in(1,1+\kappa)$ is monotonically decreasing and has the following limiting behavior
\begin{align*}
    \varphi(x)
    &=1+\kappa-\frac{\kappa}{1+\kappa}x+O(x^2)\quad\text{as }x\to0+,\\
    \varphi(x)
    &=1+\kappa/x+O(1/x^{2})\quad\text{as }x\to\infty.
\end{align*}
The rates $V$ and $\overline{V}$ are plotted in Figure~\ref{fig:mm}A. The relative difference,
\begin{align*}
    \frac{\overline{V}-V}{V},
\end{align*}
is plotted in Figure~\ref{fig:mm}B.

Comparing the well-mixed reaction rate $\overline{V}$ in \eqref{eq:mmwm} with the spatial reaction rate $V$ in \eqref{eq:mmspatial}, we make three observations. First, the well-mixed reaction rate is an overestimate,
\begin{align}\label{eq:over}
    V<\overline{V}\quad\text{for all }S/K>0\text{ and }\kappa>0.
\end{align}
Second, $V$ and $\overline{V}$ agree in the large substrate concentration regime or small trapping rate regime,
\begin{align}\label{eq:agree}
    V
    \approx\overline{V}
\quad\text{if }S/{K}\gg\kappa\text{ or }\kappa\ll1.
\end{align}
Third, $V$ and $\overline{V}$ strongly differ if $S/{K}\not\gg\kappa\gg1$. In particular, the relative difference approaches its maximum value $\kappa$ in the low substrate concentration limit,
\begin{align*}
    \lim_{S/{K}\to0}\frac{\overline{V}-V}{V}
    =\kappa,
\end{align*}
and we have that
\begin{align}\label{eq:differ}
    V
    \ll\overline{V}\quad\text{if }
    S/{K}\not\gg\kappa\gg1.
\end{align}

The three observations \eqref{eq:over}-\eqref{eq:differ} can be understood intuitively. The inequality in \eqref{eq:over} reflects the fact that $\overline{V}$ assumes that receptors are evenly distributed throughout the entire domain, which enhances the rate in which they contact diffusing substrate molecules. The approximate equality in \eqref{eq:agree} reflects the fact that (i) if $\kappa\ll1$, then receptors are sufficiently distanced from each other on the cell surface so that their restriction to the cell surface only slightly reduces the rate in which they contact diffusing substrate molecules compared to receptors being evenly distributed throughout the entire domain, and (ii) if $S/{K}\gg\kappa\not\ll1$, then catalysis is the rate limiting step and thus both $V$ and $\overline{V}$ approach the maximum possible reaction rate $V_{\max}$. Finally, \eqref{eq:differ} can be understood by noting first that if $S/{K}$ is not sufficiently large, then the arrival of substrate molecules to receptors is a rate limiting step. Further, if $\kappa\gg1$, then receptors on the same cell compete for substrate molecules, which reduces the arrival rate of substrates to receptors compared to the case where the receptors are evenly distributed throughout the entire domain.

We illustrate \eqref{eq:differ} in Figure~\ref{fig:mm}B for biophysically relevant parameter ranges. Following \cite{berg1977, wagner2006}, we take the radius of each receptor to be $\eps R=10^{-3}\,\mu\text{m}$ and the cell radius to be $R=1\,\mu\text{m}$, which is consistent with a bacterial cell or a small eukaryotic cell. Hence, $\eps=10^{-3}$. The number of receptors ${N}$ per cell can vary from roughly $N=10^2$ to $N=10^5$  \cite{perelson1997, ismael2016, lawley2020prl}, and we take ${N}=10^{4}$. Hence, the value of $\kappa$ is rather large,
\begin{align*}
    \kappa
    =\frac{\eps N}{\pi}
    \approx3.2,
\end{align*}
even though the receptors occupy only $\eps^2 N/4=2.5\times10^{-3}=0.25\%$ of the cell surface. We take the diffusion coefficient to be $D=5\times10^{2}\,\mu\text{m}^{2}\text{s}^{-1}$, which roughly corresponds to glucose uptake by an \emph{E.~coli} cell or a yeast cell \cite{meijer1996, maier2002, natarajan1999, lavrentovich2013} and chemotaxis by bacterial cells and slime mold \cite{berg1977}. The catalysis rate $\kc$ varies greatly in different biophysical systems, from around $\kc=10\,\text{s}^{-1}$ to up to $\kc=10^5\,\text{s}^{-1}$ \cite{bionumbers}. For $\kc\ge10^{2}\,\text{s}^{-1}$, Figure~\ref{fig:mm}B shows that the well-mixed estimate $\overline{V}$ can significantly overestimate $V$ for substrate concentrations $S$ less than about $1\,\mu\text{M}$.

%%%%%%%%%%%%%%%%%%%%%%%%%%%%%%%%%%%%%%%%%%%%%%%%%%%%%%%%%%%%%%%%%%%%%%%%%%%%%%%%%%%%%%%%%%%%%%%%%%%%%%%%%%%%
\subsection{Substrate Competition\label{sec:sc}}

Consider the Michaelis-Menten reaction scheme in \eqref{eq:mm000}, but now suppose that there are two substrates which compete for the enzymes/receptors,
\begin{align}\label{eq:sc}
\begin{split}
\schemestart
\chemfig{S_1} + E \arrow{->[$k_{\textup{a},1}$]} \chemfig{ES_1} \arrow{->[$k_{\textup{c},1}$]} E + \chemfig{P},
\schemestop\\
\schemestart
\chemfig{{S_2}} + E \arrow{->[$k_{\textup{a},2}$]} \chemfig{E{S_2}} \arrow{->[$k_{\textup{c},2}$]} E + \chemfig{P}.
\schemestop 
\end{split}
\end{align}
The well-mixed reaction rate $\overline{V}$ for \eqref{eq:sc000} is \cite{pocklington_1969}
\begin{equation}\label{eq:sc000}
\overline{V}
=\frac{{V}_{\max,1}S_1}{{K_1} + ({K_1}/{K_2}){S_2}+S_1} + \frac{{V}_{\max,2}{S_2}}{{K_2} + ({K_2}/{K_1})S_1 + S_2},
\end{equation}
where $S_1$ and ${S_2}$ are the substrate concentrations of species 1 and 2 and
\begin{align*}
    V_{\max,1}
    &=\kca E_{\text{tot}},\\
    V_{\max,2}
    &=\kcb E_{\text{tot}},\\
    {K_1}
    &=\kca/\kaa,\\
    {K_2}
    &=\kcb/\kab.
\end{align*}
In the appendix, we derive the following reaction rate for the corresponding spatial system,
\begin{align}\label{eq:scspatial}
\begin{split}
    V
    &=\frac{{V}_{\max,1}S_1}{\varphi(S_1/{K_1}+S_2/{K_2}){K_1} + ({K_1}/{K_2}){S_2} + S_1}\\
    &\quad + \frac{{V}_{\max,2}{S_2}}{\varphi(S_1/{K_1}+S_2/{K_2}) {K_2} + ({K_2}/{K_1})S_1 + S_2},
\end{split}
\end{align}
where $\varphi(x)$ is the dimensionless function in \eqref{eq:varphi}.

Analogous to \eqref{eq:over}-\eqref{eq:differ} for Michaelis-Menten kinetics, comparing the well-mixed reaction rate $\overline{V}$ in \eqref{eq:sc} and the spatial reaction rate $V$ in \eqref{eq:scspatial} shows that for the Substrate Competition kinetics in \eqref{eq:sc000},
\begin{align*}
    V
    &<\overline{V}\quad\text{for all }S_1/K_1>0,\,S_2/K_2>0,\text{ and }\kappa>0,\\
    V
    &\approx \overline{V}\quad\text{if }S_1/K_1+S_2/K_2\gg\kappa\text{ or }\kappa\ll1,\\
    V
    &\ll \overline{V}\quad\text{if }
    S_1/K_1+S_2/K_2\not\gg\kappa\gg1.
\end{align*}

%%%%%%%%%%%%%%%%%%%%%%%%%%%%%%%%%%%%%%%%%%%%%%%%%%%%%%%%%%%%%%%%%%%%%%%%%%%%%%%%%%%%%%%%%%%%%%%%%%%%%%%%%%%%
\subsection{Competitive Inhibition\label{sec:ci}}

In addition to the Michaelis-Menten reaction in \eqref{eq:mm000}, suppose the enzyme/receptor E can be bound by an inhibitor I,
\begin{align}\label{eq:ci000}
\begin{split}
    \schemestart 
    S + E \arrow{->[$\ka$]} ES \arrow{->[$\kc$]} E + P 
\schemestop \\
\schemestart
    I + E \arrow {<=>[$\ki$][$\kd$]} EI. \quad\qquad\qquad\qquad\phantom{a}
\schemestop
\end{split}
\end{align}
The well-mixed reaction rate $\overline{V}$ for the so-called Competitive Inhibition in \eqref{eq:ci000} is \cite{dixon_1953}
\begin{align}\label{eq:ciwm}
{\overline{V}}
=\frac{{V}_{\max}S}{{K}+({K}/{K_I})I+S},
\end{align}
where $S$ is the substrate concentration, $I$ is the inhibitor concentration, $V_{\max}$ and ${K}$ are in \eqref{eq:mmend}, and 
\begin{align}\label{eq:ciend}
    {K_I}
    &=\kd/\ki.
\end{align}

\begin{figure}[t]
\centering
\includegraphics[width=1\textwidth]
{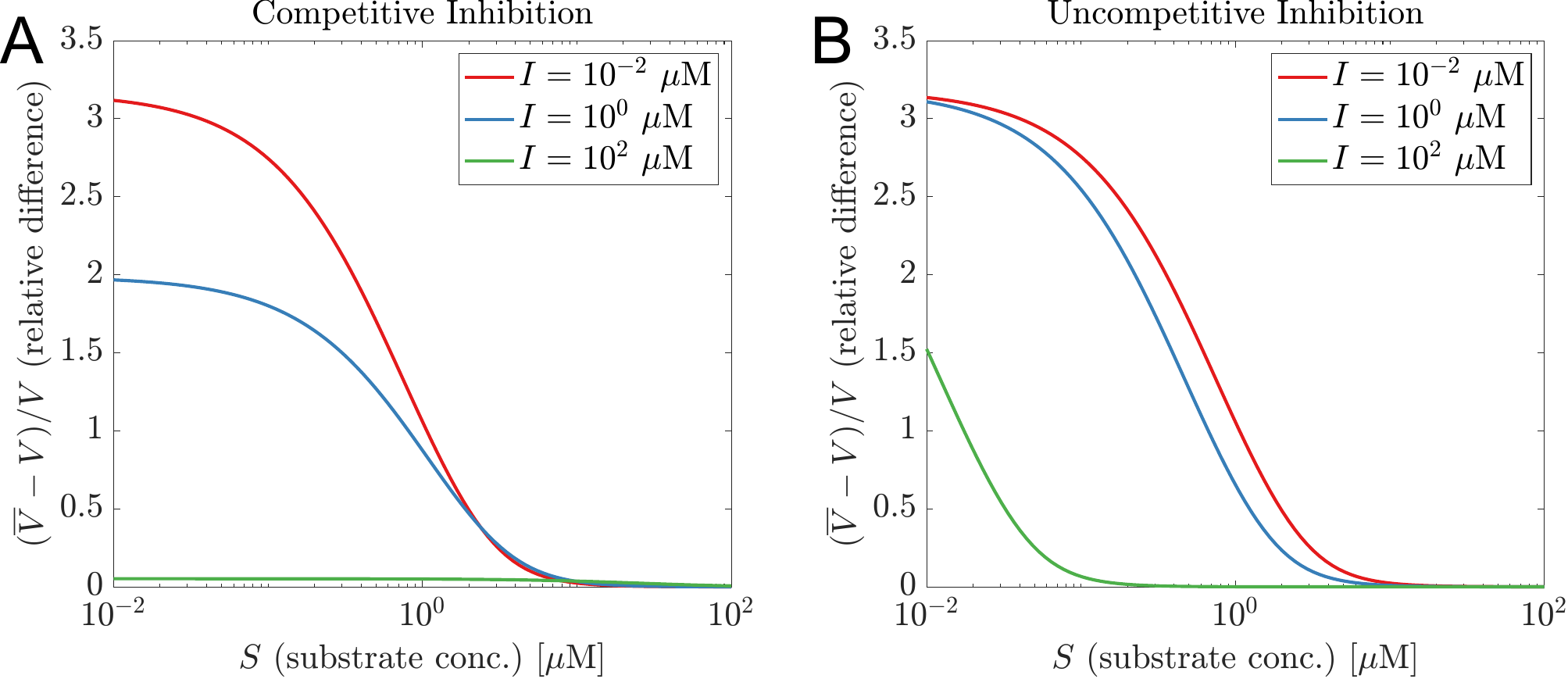}
\caption{\label{fig:ciui} The effects of inhibition. Panel A: For Competitive Inhibition described in section~\ref{sec:ci}, the relative difference between the spatial reaction rate $V$ in \eqref{eq:cispatial} and the well-mixed reaction rate $\overline{V}$ in \eqref{eq:ciwm} increases at low substrate concentrations $S$ and low inhibitor concentrations $I$. Panel B: For Uncompetitive Inhibition described in section~\ref{sec:ui}, the relative difference between the spatial reaction rate $V$ in \eqref{eq:uispatial} and the well-mixed reaction rate $\overline{V}$ in \eqref{eq:ui} increases at low substrate concentrations $S$ and low inhibitor concentrations $I$. In both panels, we take $\kc=10^3\,/\text{s}$, $D_{\textup{i}}$, and the other parameters are given in section~\ref{sec:mm}.} 
\end{figure}

In the appendix, we derive the following reaction rate for the corresponding spatial system,
\begin{align}\label{eq:cispatial}
    V
    &=\frac{{V}_{\max}S}{\psi(S/{K},I/{K_I}){K}+({K}/{K_I})I+S},
\end{align}
where $\ki=4\eps D_{\textup{i}} R$ with $D_{\textup{i}}$ the inhibitor diffusivity (analogous to $\ka$ in \eqref{eq:ka}) and $\psi(x,y)$ is the following dimensionless function,
\begin{align*}
    \psi(x,y)
    =&\frac{1}{2}\Big(1+\kappa-x-y+\sqrt{\big(1+\kappa+x+y\big)^2-4\kappa x}\Big).
\end{align*}
Note that $\psi(x,y)\in(1,1+\kappa)$ is a decreasing function of $x>0$, an increasing function of $y>0$, and has the following limiting behavior,
\begin{align*}
    \psi(x,y)
    &=1+\kappa-\frac{\kappa}{1+\kappa+y}x+O(x^2)\quad\text{as }x\to0+,\\
    \psi(x,y)
    &=1+\kappa(1+y)/x+O(1/x^2)\quad\text{as }x\to\infty,\\
    \psi(x,y)
    &=\varphi(x)+O(y)\quad\text{as }y\to0+,\\
    \psi(x,y)
    &=1+\kappa-\kappa x/y+O(1/y^2)\quad\text{as }y\to\infty.
\end{align*}
It follows that for the Competitive Inhibition kinetics in \eqref{eq:ci000},
\begin{align}
    V
    &<\overline{V}\quad\text{for all }S/K>0,\,I/K_I>0,\text{ and }\kappa>0,\nonumber\\
    V
    &\approx \overline{V}\quad\text{if }S/K\gg\kappa(1+I/K_I)\text{ or }\kappa\ll1,\nonumber\\
    V
    &\ll \overline{V}\quad\text{if }
    S/K\not\gg\kappa(1+I/K_I)\text{ and }\kappa\gg1.\label{eq:cimuchless}
\end{align}
Comparing \eqref{eq:cimuchless} with \eqref{eq:differ}, observe that the presence of a competitive inhibitor shrinks the region of parameter space in which the well-mixed reaction rate $\overline{V}$ overestimates the spatial reaction rate $V$ (see Figure~\ref{fig:ciui}A).

%%%%%%%%%%%%%%%%%%%%%%%%%%%%%%%%%%%%%%%%%%%%%%%%%%%%%%%%%%%%%%%%%%%%%%%%%%%%%%%%%%%%%%%%%%%%%%%%%%%%%%%%%%%%
\subsection{Uncompetitive Inhibition\label{sec:ui}}

Rather than the Competitive Inhibition in \eqref{eq:mm000}, suppose that the complex ES can be bound by an inhibitor I,
\begin{align}\label{eq:ui000}
\begin{split}
    \schemestart
S\phantom{E} + E \arrow {->[$\ka$]} ES \arrow {->[$\kc$]} E + P 
\schemestop \\
\schemestart 
ES + I \arrow {<=>[$\ki$][$\kd$]} ESI\quad\qquad\qquad\qquad\phantom{ab}
\schemestop
\end{split}
\end{align}
The well-mixed reaction rate $\overline{V}$ for the so-called Uncompetitive Inhibition in \eqref{eq:ui000} is \cite{dixon_1953}
\begin{align}\label{eq:ui}
{\overline{V}}
=\frac{{V}_{\max}S}{{K} + (1+I/{K_I})S},
\end{align}
where $S$ is the substrate concentration, $I$ is the inhibitor concentration, $V_{\max}$ and ${K}$ are in \eqref{eq:mmend}, and ${K_I}$ is in \eqref{eq:ciend}.

In the appendix, we derive the following reaction rate for the corresponding spatial system,
\begin{align}\label{eq:uispatial}
    V
    =\frac{{V}_{\max}S}{\varphi((1+I/{K_I})S/{K}){K} + (1+I/{K_I})S},
\end{align}
where $\varphi(x)$ is the dimensionless function in \eqref{eq:varphi}.

Analogous to \eqref{eq:over}-\eqref{eq:differ} for Michaelis-Menten kinetics, comparing the well-mixed reaction rate $\overline{V}$ in \eqref{eq:ui} and the spatial reaction rate $V$ in \eqref{eq:uispatial} shows that for the Uncompetitive Inhibition kinetics in \eqref{eq:ui000},
\begin{align}
    V
    &<\overline{V}\quad\text{for all }S/K>0,\,I/K_I>0,\text{ and }\kappa>0,\nonumber\\
    V
    &\approx \overline{V}\quad\text{if }
    (1+I/K_I)S/K\gg\kappa\text{ or }\kappa\ll1,\nonumber\\
    V
    &\ll \overline{V}\quad\text{if }
    (1+I/K_I)S/K\not\gg\kappa\gg1.\label{eq:uimuchhless}
\end{align}
Comparing \eqref{eq:uimuchhless} and \eqref{eq:differ}, observe that the presence of an uncompetitive inhibitor decreases the region of parameter space in which $V\ll\overline{V}$ (see Figure~\ref{fig:ciui}B).

%%%%%%%%%%%%%%%%%%%%%%%%%%%%%%%%%%%%%%%%%%%%%%%%%%%%%%%%%%%%%%%%%%%%%%%%%%%%%%%%%%%%%%%%%%%%%%%%%%%%%%%%%%%%%%%%%%%%%%%%%%%%%%%%%%%%%%%%%%%%%%%%%%%%%%%%%%%%%%%%%%%%%%%%%%%%%%%%%%%%%%%%%%%%%%%%%%%%%%%%%%%%%%%%%%%%%%%%%%%%%%%%%%%%%%%%%%%%%%%%%%%%%%%%%%%%%%%%%%%%%%%%%%%%%%%%%%%%%%%%%%%%%%%%%%%%%%%%%%%%%%%%%%%%%%%%%%%%%%%%%%%%%%%%%%%%%%%%%%%%%%%%%%%%%%%%%%%%%%%%%%%%%%%%%%%%%%%%%%%%%%%%%%%%%%%%%%%%%%%%%%%%%%%%%%%%%%%%%%%%%%%%%%%%%%%%%%%%%%%%%%%%%%%%%%%%%%%%%%%%%%%%%%%%%%%%%%%%%%%%%%%%%%%%%%%%%%%%%%%%%%%%%%%%%%%%%%%%%%%%%%%%%%%%%%%%%%%%%%%%%%%%%%%%%%%%%%%%%%%%%%%%%%%%%%%%%%%%%%%%%%%%%%%%%%%%%%%%%%%%%%%%%%%%%%%%%%%%%%%%%%%%%%%%%%%%%%%%%%%%%%%%%%%%
\section{Conclusion\label{sec:conc}}

In this work, we developed and analyzed a spatial model to estimate the influx rate of diffusing molecules through membrane-bound receptors. We analyzed several prototypical interaction schemes and compared our spatially-resolved estimates to classical well-mixed theories. Our results predict biophysical parameter regimes in which spatial features strongly affect kinetics, and we proposed how to modify the well-mixed estimates in such regimes.

From a mathematical perspective, our models employed boundary homogenization theory to couple a PDE describing bulk diffusion to nonlinear ODEs describing reactions on a lower-dimensional surface. While we analyzed our system at steady-state, several other interesting works have found that similarly coupled PDE-ODE systems can exhibit rich temporal dynamics \cite{levine2005, gomez2007, gou2016, gou2017, gomez2019, david2020, pb6, gomez2021pattern}. In our analysis, we assumed that different cells were sufficiently separated, which allowed us to analyze each cell in isolation. Relaxing this assumption would significantly complicate the mathematical analysis, though strong localized perturbation theory \cite{gomez2021pattern} has been used to study similar systems. 

From a biological perspective, our analysis highlights how spatial features can affect reaction timescales in cell biology. In particular, naively applying well-mixed theories which ignore spatial inhomogeneities can sometimes lead to false conclusions. For instance, as pointed out in \cite{ma2020strong}, mathematical models which resolve spatial features have yielded important insights into general protein kinase signaling, cyclic AMP signaling, T cell synapse formation, and B cell activation. From a modeling perspective, an important avenue for future would is to develop a more general understanding (or perhaps merely rules of thumb) of when spatial details can or cannot be safely ignored when building models.

Finally, our results can be interpreted as a theoretical derivation of the Monod equation. The Monod equation is used to describe the relationship between bacterial growth and nutrient uptake \cite{reardon_2000} and takes the form of the well-mixed Michaelis-Menten reaction rate in \eqref{eq:mm_rex_rate0}. Though empirically verified in many studies, the Monod equation long escaped mechanistic justification \cite{liu2007overview}. Our results for Michaelis-Menten interactions can model the transport of diffusing nutrients into bacteria and provide an alternative for recent mechanistic derivations of the Monod equation \cite{enouy_2022, alvarez_2019}. Our relatively simple derivation emphasizes the role played by spatial and biophysical factors and provides a framework to study microbial growth in a wide array of nutrient uptake scenarios. 

\appendix
\section{Appendix\label{sec:appendix}}

%%%%%%%%%%%%%%%%%%%%%%%%%%%%%%%%%%%%%%%%%%%%%%%%%%%%%%%%%%%%%%%%%%%%%%%%%%%%%%%%%%%%%%%%%%%%%%%%%%%%%%%%%%%%%%%%%%%%%%%%%%%%%%%%%%%%%%%%%%%%%%%%%%%%%%%%%%%%%%%%%%%%%%%%%%%%%%%%%%%%

%%%%%%%%%%%%%%%%%%%%%%%%%%%%%%%%%%%%%%%%%%%%%%%%%%%%%%%%%%%%%%%%%%%%%%%%%%%%%%%%%%%%%%%%%%%%%%%%%%%%%%%%%%%%%%%%%%%%%%%%%%%%%%%%%%%%%%%%%%%%%%%%%%%%%%%%%%%%%%%%%%%%%%%%%%%%%%%%%%%%
\subsection{Uptake derivation for Substrate Competition}
Here, we describe the reaction kinetics of membrane receptors that feature Substrate Competition shown in Figure \ref{fig:schemes}b and derive the molecular influx for such systems. 

Consider a spherical cell $\Omega$ of radius $R$ centered at the origin with two species of diffusing substrate molecules surrounding it and denote their concentrations by $c_1=c_1(r,\theta,\phi,t)$ and $c_2=c_2(r,\theta,\phi,t)$. Assume that the concentrations of the two species of diffusing molecules away from the cell are fixed at $\cfara>0$ and $\cfarb>0$ (this reflects our assumption that different cells are sufficiently well-separated that they can be treated as non-interacting). Then, the concentrations $c_1$ and $c_2$ satisfy the following diffusion PDEs and far-field conditions:
\begin{align}\begin{split}\partial_t c_1 &= D_1 \Delta c_1, \quad r>R, \\
\partial_t c_2 &= D_2 \Delta c_2, \quad r>R, \\
\lim_{r\to\infty} c_1&=\cfara>0,\\
\lim_{r\to\infty} c_2&=\cfarb>0,\end{split}\label{eq:sc_pde}\end{align}
where $D_1$ and $D_2$ are the diffusivities of the two molecular compounds. In this interaction scenario, we assume that the receptors, represented by E, have one active site for the competing diffusing molecules, represented by $\text{S}_1$ and $\text{S}_2$. Thus, the diffusing molecules can form two complexes with the receptors, a species-1-bound receptor complex $\text{ES}_1$, and a species-2-bound receptor complex $\text{ES}_2$. These complexes can be ``catalyzed'' to yield the product P. Based on this scheme, the law of mass action can be used to describe how the concentration of the diffusing molecules changes on the boundary,
\begin{align}\begin{split}D_1\partial_r c_1&=\kaa c_1 u, \quad r=R, \\
D_2\partial_r c_2&=\kab c_2 u, \quad r=R,\end{split}\label{eq:sc_bc}\end{align}
where the concentration of available receptors, species-1-bound receptor complexes, and species-2-bound receptor complexes at time $t$ are respectively denoted by $u(t)$, $b_1(t)$, and $b_2(t)$ and governed by
\[\frac{\dd}{\dd t}u&=-\kaa c_1(R) u - \kab c_2(R) u + b_1 \kca + b_2 \kcb, \\
\frac{\dd}{\dd t}b_1&=\kaa c_1(R) u -b_1\kca, \\
\frac{\dd}{\dd t}b_2&=\kab c_2(R) u - b_2\kcb,\]
with $c_1(R):=c_1(R,\theta,\phi,t)$ and $c_2(R):=c_2(R,\theta,\phi,t)$ for notational ease. In this system, receptors can be in one of three states: available, part of a species-1-bound complex, or part of a species-2-bound complex. Therefore, the total receptor concentration in this system is conserved. Since we have $N$ receptors, the conserved total receptor concentration is
\begin{align*}
    u_0:=u(t)+b_1(t)+b_2(t)=\frac{N}{4\pi R^2}.
\end{align*}
This conservation law can then be used to reduce the system of three ODEs to a system of two ODEs and create the following system of two linear algebraic equations that describe the steady-state behavior of this system, 
\[0=\kaa \cinfa(R) (u_0-\binfa-\binfb) -\binfa\kca, \\
0=\kab \cinfb(R) (u_0-\binfa-\binfb)-\binfb\kcb,\]
where $\cinfa, \cinfb, \binfa$, and $\binfb,$ denote the steady-state concentrations of the diffusing molecules of
species 1, of species 2, and the steady-state concentration of the species-1-bound and the species-2-bound receptor complexes, respectively. The solution to this linear system is 
\[\binfa=\frac{\cinfa(R)\kaa\kcb u_0}{\cinfa(R)\kaa\kcb+\cinfb(R)\kab\kca+\kca\kcb},\\
\binfb=\frac{\cinfb(R)\kab\kca u_0}{\cinfa(R)\kaa\kcb+\cinfb(R)\kab\kca+\kca\kcb}.\]

In steady-state, the solutions to the PDEs in \eqref{eq:sc_pde} evaluated at $r=R$ are  
\begin{align}
\begin{split}\cinfa(R)&=\cfara\left(1-a_1\right),\\
\cinfb(R)&=\cfarb\left(1-a_2\right), \end{split}
\label{eq:cinf_sc}
\end{align}
with $a_1$ and $a_2$ as undetermined integration constants. Substituting $\cinfa(R)$ and $\cinfb(R)$ into the boundary conditions in \eqref{eq:sc_bc} given that $\kaa=4\eps D_1 R$ and $\kab=4\eps D_2 R$, $a_1$ can be shown to satisfy $a_1=a_2$ and to be a root of the following quadratic polynomial in $a_1$,
\[A_2a_1^2+A_1a_1+A_0,\]
if we define
\[A_2
&=\cfara/{K_1} + \cfarb/{K_2}\\
A_1&=-(\kappa + \cfara/{K_1} +\cfarb/{K_2} + 1)\\
A_0&=\kappa.\]

Using the quadratic formula and the condition that $0< a_1 <1$ due to steady-state concentration in \eqref{eq:cinf_sc} being positive, 
the solution to the above polynomial is given by
\begin{align}a_1=\frac{-A_1 - \sqrt{A_1^2 - 4 A_2A_0}}{2A_2}=\frac{-2A_0}{A_1-\sqrt{A_1^2-4A_2A_0}}. \label{eq:quad_formsc}
\end{align}
We compute the reaction rate $V$ by multiplying the cell concentration $C$ by the molecular influx into a single cell,
\begin{align}\begin{split}
V
&:=C\int_{\Omega} D_1 \nabla c_1 \cdot \, d\sigma + \int_{\Omega} D_2 \nabla c_2 \cdot \, d\sigma \\
&=C4\pi R (D_1\cfara+D_2\cfarb)a_1.
\end{split}
\end{align}
Plugging back $A_0, A_1$, and $A_2$ and $u_0=N/(4\pi R^2)$, the reaction rate $V$ can be algebraically manipulated to have the form in \eqref{eq:scspatial}.

\subsection{Uptake derivation for Competitive Inhibition \label{subsec:comp_inh}}
Here, we describe the reaction kinetics of membrane receptors that feature Competitive Inhibition shown in Figure \ref{fig:schemes}c and derive the molecular influx for such systems. 

Consider a spherical cell $\Omega$ of radius $R$ centered at the origin with one species of diffusing substrate molecules and one species of diffusing inhibitor molecules surrounding it, and denote their concentrations by $c=c(r,\theta,\phi,t)$ and $m=m(r,\theta,\phi,t)$. Assume that the concentrations of the two species of diffusing molecules away from the cell are fixed at $\cfar>0$ and $\mfar>0$ (this reflects our assumption that different cells are sufficiently well-separated that they can be treated as non-interacting). Then, the concentrations $c$ and $m$ satisfy the following diffusion PDEs and far-field conditions,
\begin{align}\begin{split}\partial_t c &= {{D}} \Delta c, \quad r>R, \\
\partial_t m &= D_{\textup{i}} \Delta m, \quad r>R, \\
\lim_{r\to\infty} c&=\cfar>0,\\
\lim_{r\to\infty} m&=\mfar>0,\end{split}\label{eq:comp_inh_pde}\end{align}
where ${{D}}$ and $D_{\textup{i}}$ are the substrate and inhibitor diffusivities, respectively. In this interaction scenario, we assume that the receptors, represented by E, have one active site for the competing diffusing substrate and inhibitor molecules, represented by S and I. Thus, the diffusing molecules can form two complexes with receptors, the substrate-bound complex ES and an inhibitor-bound complex EI. Only the substrate-bound complex ES can be ``catalyzed'' to yield the product P. The inhibitor binding is reversible, and thus EI can dissociate at rate $\kd$. Based on this scheme, the law of mass action can be used to describe how the concentration of the diffusing molecules changes on the boundary,
\begin{align}\begin{split}{{D}} \partial_r c&=\ka c u, \quad r=R, \\
D_{\textup{i}}\partial_r m&=\ki mu -\kd b_m, \quad r=R, \end{split}\label{eq:comp_inh_bc}\end{align}
where the concentration of available receptors, substrate-bound receptor complex, and the inhibitor-bound receptor complex at time $t$ are respectively denoted by $u(t), b_c(t)$, and $b_m(t)$ and governed by 
\[\frac{\dd}{\dd t}u&=-\ka c(R)u+\kc b_c - \ki m(R) u + \kd b_m, \\
\frac{\dd}{\dd t}b_c&=\ka c(R) u -\kc b_c, \\
\frac{\dd}{\dd t}b_m&=\ki m(R)u - \kd b_m, \]
with $c(R):=c(R,\theta,\phi,t)$ and $m(R):=m(R,\theta,\phi,t)$ for notational ease.  In this system, receptors can be in one of
three states: available, part of a substrate-bound complex,
or part of an inhibitor-bound complex. Therefore, the total
receptor concentration in this system is conserved. Since we
have $N$ receptors, the conserved total receptor
concentration is $u_0:=u(t)+b_c(t)+b_m(t)=\frac{N}{4\pi R^2}$. The
conservation law can then be used to reduce the system of three ODEs to a system of two ODEs and create the following system of two linear algebraic equations that describe the steady-state
behavior of this system,
\[0&=\ka \cinf(R) (u_0-\binfc -\binfm )-\kc \binfc, \\
0&=\ki \minf(R) (u_0-\binfc -\binfm )-\kd \binfm, \]
where $\cinf,\minf$ and $\binfc,\binfm$ denote the steady-state concentration of the substrate molecules, inhibitor molecules, substrate-bound complex, and the inhibitor-bound complex, respectively. The solution to this linear system is 
\[\binfc&=\frac{\cinf(R)\ka\kd u_0}{\cinf(R) \ka \kd + \minf(R)\kc \ki+ \kc \kd },\\
\binfm&=\frac{\minf(R) \kc \ki u_0}{\cinf(R) \ka \kd + \minf(R)\kc \ki + \kc \kd }.\]

In steady-state, the solutions to the PDEs in \eqref{eq:comp_inh_pde} evaluated at $r = R$ are
\begin{align}\begin{split}\cinf(R)&=\cfar \left(1-a_1\right),\\
\minf(R)&=\mfar \left(1-a_2\right),\end{split}\label{eq:comp_inh_cinf_minf}\end{align}
with $a_1$ and $a_2$ as undetermined integration constants. Substituting $\cinf(R)$ and $\minf(R)$ into the boundary conditions in \eqref{eq:comp_inh_bc} given that $\ka=4\eps {{D}} R$ and $\ki=4\eps D_{\textup{i}} R$, $a_2$ can be shown to satisfy $a_2=0$ and $a_1$ can be shown to be a root of the following quadratic polynomial in $a_1$, 
\begin{align}
A_2 a_1^2+ A_1 a_1 + A_0, \label{eq:comp_inh_poly}\end{align}
if we define ${K_I}=\kd/\ki$ and 
\[A_2&:=\cfar/{K}, \\
A_1&:=-(\kappa +\cfar/{K} + \mfar/{K_I}  + 1 ),
\\
A_0&:=\kappa. \]

Using the quadratic formula and the condition that $0< a_1 <1$ due to steady-state concentration in Eq.\eqref{eq:comp_inh_cinf_minf} being positive, 
the solution to the above polynomial is given in \eqref{eq:quad_form}. 
We compute the reaction rate $V$ by multiplying the cell concentration $C$ by the molecular influx into a single cell,
\begin{align}\begin{split}V&:=C\int_{\Omega} {{D}} \nabla c \cdot d\sigma\\
&=C4\pi R {{D}} \cfar a_1.
\end{split}\label{eq:J_inh}\end{align}
Plugging back $A_0, A_1$ and $A_2$ and $u_0=N/(4\pi R^2)$, the reaction rate $V$ can algebraically manipulated to have the form in \eqref{eq:cispatial}.

%%%%%%%%%%%%%%%%%%%%%%%%%%%%%%%%%%%%%%%%%%%%%%%%%%%%%%%%%%%%%%%%%%%%%%%%%%%%%%%%%%%%%%%%%%%%%%%%%%%%%%%%%%%%
\subsection{Uptake derivation for Uncompetitive Inhibition}
Here, we describe the reaction kinetics of membrane receptors that feature Uncompetitive Inhibition shown in Figure \ref{fig:schemes}d and derive the molecular influx for such systems. 

Consider a spherical cell $\Omega$ and diffusing substrate and inhibitor species as done in subsection~\ref{subsec:comp_inh} and the PDE problem in ~\eqref{eq:comp_inh_pde}. In this interaction scenario, we assume that the receptors, represented by E, have one active site for the diffusing substrate molecules, represented by S. Thus, the diffusing molecules can form the substrate-bound complex ES. This ES complex can either be ``catalyzed'' to yield the product P or reversibly bind with the diffusing inhibitor species and form the receptor-substrate-inhibitor complex, ESI, which is an inactive state unable to be catalyzed. Based on this scheme, the law of mass action can be used to describe how the concentration of the diffusing molecules changes on the boundary:  
\begin{align}\begin{split}{{D}} \partial_r c&=\ka c u, \quad r=R, \\
D_{\textup{i}}\partial_r m&=\ki m b_c -\kd b_m, \quad r=R, \end{split} \label{eq:comp_uninh_bc}\end{align}
where the concentration of available receptors, substrate-bound receptor complex, and the inhibitor-substrate-bound receptor complex at time $t$ are respectively denoted by $u(t), b_c(t)$, and $b_m(t)$ and governed by 
\[\frac{\dd}{\dd t}u&=-\ka c(R)u+\kc b_c,  \\
\frac{\dd}{\dd t}b_c&=\ka c(R) u -\kc b_c -\ki m(R) b_c + \kd b_m, \\
\frac{\dd}{\dd t}b_m&=\ki m(R) b_c - \kd b_m, \]
with $c(R):=c(R,\theta,\phi,t)$ and $m(R):=m(R,\theta,\phi,t)$ for notational ease. In this system, receptors can be in one of three states: available, part of a substrate-bound complex, or part of an inhibitor-substrate-bound complex. Therefore, the total receptor concentration in this system is conserved. Since we have $N$ receptors, the conserved total receptor concentration is $u_0:=u(t)+b_c(t)+b_m(t)=\frac{N}{4\pi R^2}$. The conservation law can then be used to reduce the system of three ODEs to a system of two ODEs and create the following system of two linear algebraic equations that describe the steady-state behavior of the system,
\[0&=\ka \cinf(R) \uinf -\kc (u_0-\uinf-\binfm), \\
0&=\ki \minf(R) (u_0-\uinf -\binfm )-\kd \binfm, \]
where $\cinf, \minf$ and $\uinf, \binfm$ denote the steady-state concentration of the substrate molecules, inhibitor molecules, available receptors, and the inhibitor-substrate-bound complex, respectively. The solution to this linear system is  
\[\uinf &=\frac{\kc\kd u_0}{\cinf(R) \ka \ki \minf(R)+\cinf(R) \ka \kd + \kc \kd},\\
\binfm&=\frac{u_0\cinf(R) \ka\ki \minf(R)}{\cinf(R) \ka \ki \minf(R)+\cinf(R) \ka \kd + \kc \kd}.\]

In steady-state, the solutions to the PDEs in \eqref{eq:comp_inh_pde} evaluated at $r=R$, $\cinf(R)$ and $\minf(R)$ respectively, are given in \eqref{eq:comp_inh_cinf_minf} with $a_1$ and $a_2$ as undetermined integration constants. Substituting $\cinf$ and $\minf$ into the boundary conditions in \eqref{eq:comp_uninh_bc} given that $\ka=4\eps {{D}} R$ and $\ki=4 \eps D_{\textup{i}} R$, $a_2$ can be shown to satisfy $a_2=0$, and $a_1$ can be shown to be a root of the following quadratic polynomial in $a_1$, 
\begin{align}\begin{split}
A_{2} a_1^2 + A_1a_1 +A_0, \end{split}\label{eq:uncomp_inh_poly}\end{align}
if the following new variables are defined for convenience: 
\[A_{2}&=(S/{K})(I/{K_I})  +S/{K}, \\
A_{1}&=- ((S/{K})(I/{K_I}) + S/{K} +\kappa+ 1 ), \\
A_0&=\kappa.\]
Then $a_1$ is given by \eqref{eq:quad_form} and the molecular influx $J$ is given by \eqref{eq:J_inh}. Plugging back $A_0, A_1$ and $A_2$ and $u_0=N/(4\pi R^2)$, the reaction rate $V$ can be algebraically manipulated to have the form in \eqref{eq:uispatial}.

%%%%%%%%%%%%%%%%%%%%%%%%%%%%%%%%%%%%%%%%%%%%%%%%%%%%%%%%%%%%%%%%%%%%%%%%%%%%%%%%%%%%%%%%%%%%%%%%%%%%%%%%%%%%%%%%%%%%%%%%%%%%%%%%%%%%%%%%%%%%
\subsubsection*{Acknowledgments}
The authors were supported by the National Science Foundation (Grant Nos.\ CAREER DMS-1944574 and DMS-2325258).

\subsection*{Data availability statement}

This manuscript has no associated data.

%%%%%%%%%%%%%%%%%%%%%%%%%%%%%%%%%%%%%%%%%%%%%%%%%%%%%%%%%%%%%%%%%%%%%%%%%%%%%%%%%%%%%%%%%%%%%%%%%%%%%%%%%%%%%%%%%%%%%%%%%%%%%%%%%%%%%%%%%%%%

\bibliography{mybib.bib}
\bibliographystyle{unsrt}

%%%%%%%%%%%%%%%%%%%%%%%%%%%%%%%%%%%%%%%%%%%%%%%%%%%%%%%%%%%%%%%%%%%%%%%%%%%%%%%%%%%%%%%%%%%%%%%%%%%%%%%%%%%%%%%%%%%%%%%%%%%%%%%%%%%%%%%%%%%%%%%%%%%%%%%%%%%%%%%%%%%%%%%%%%%%%%%%%%%%%%%%%%%%%%%%%%%%%%%%%%%%%%%%%%%%%%%%%%%%%%%%
\end{document}